# Frequency-domain Parallel Computing Using Single On-Chip Nonlinear Acoustic-wave Device

Jun Ji[1,2,*], Zichen Xi[1,2], Bernadeta R. Srijanto[3], Ivan I. Kravchenko[3], Ming Jin[1], Wenjie Xiong[1], and Linbo Shao[1,2,*]

[1]Bradley Department of Electric and Computer Engineering, Virginia Tech, Blacksburg, VA USA
[2]Viginia Tech Center for Quantum Information Science and Engineering (VTQ), Blacksburg, VA USA
[3]Center of Nanophase Materials Sciences, Oak Ridge National Laboratory, Oak Ridge, TN USA
*Email: junji@vt.edu, shaolb@vt.edu

*Abstract*—Multiply-accumulation (MAC) is a crucial computing operation in signal processing, numerical simulations, and machine learning. This work presents a scalable, programmable, frequency-domain parallel computing leveraging gigahertz (GHz)-frequency acoustic-wave nonlinearities. By encoding data in the frequency domain, a single nonlinear acoustic-wave device can perform a billion arithmetic operations simultaneously. A single device with a footprint of 0.03 mm$^2$ on lithium niobate (LN) achieves 0.0144 tera floating-point operations per second (TFLOPS), leading to a computing area density of 0.48 TFLOPS/mm$^2$ and a core power efficiency of 0.14 TFLOPS/Watt. As applications, we demonstrate multiplications of two 16-by-16 matrices and convolutional imaging processing of 128-by-128-pixel photos. Our technology could find versatile applications in near-sensor signal processing and edge computing.

## I. Introduction

Compared to conventional digital computing, analog computing [1-4] provides more power-efficient solutions to size-weight-and-power (SWaP)-constrained applications, especially in near-sensor data processing and artificial neural networks. While analog computing systems can be simply scaled up by increasing the number of devices using a larger chip area, this may result in larger computing errors due to the fabrication and environment (such as temperature) variations between different devices on chip. Alternatively, synthetic frequency dimension of waves [5] can be leveraged to scale up the computing capability using a single device. Photonic integrated circuits using acousto-optic [6] or electro-optic [7] modulations demonstrate unprecedented computing capabilities, despite current challenges in fully integrating all components including laser, modulators, multiplexers, photodiodes, etc. On the other hand, GHz mechanical waves on solids can be efficiently and bidirectionally transduced to electrical signals, and their sub-micron wavelength promises small device footprints.

Here, we demonstrate parallel computing leveraging the acoustic-wave nonlinearity of LN. Vectors or matrices of data are simultaneously encoded in the frequency domain, and MAC operations are realized by the second-order nonlinear processes of our devices. Our frequency-domain computing paradigm does not rely on any device structures to predefine arithmetic operations or operands. As demonstrated in this work, our device is fully in situ programmable and can be reconfigured to perform matrix multiplications and convolutions at different scales.

## II. Device Design and Computing Principles

LN as a non-centrosymmetric material possesses strong elastic and piezoelectric nonlinearity, enabling arithmetic operations of signals encoded at different frequencies. Our frequency-domain parallel computing device is on a Z-cut LN with surface acoustic wave (SAW) propagating in $Y$ direction (Fig. 1(a)). To excite and detect SAW, 100-nm-thick aluminum interdigital transducers (IDTs) are patterned by electron beam lithography and deposited using an electron beam evaporator followed by lift-off process (Figs. 1(b)-(c)). The device consists of two types of IDTs, working at the fundamental frequency (i.e., $f_0 = 1{,}023.5$ MHz) and the second-order frequency (i.e., 2,047 MHz), respectively. The pitch of fundamental (second-order) IDT fingers $p_f$ ($p_s$) is 1.84 (0.908) µm, which corresponds to the half wavelength of SAW at the fundamental (second-order) frequency (Fig. 2).

Fig. 3 illustrates the principle of our frequency-domain nonlinear computing. Floating-point input vectors $\vec{a}$ and $\vec{b}$ are simultaneously encoded in the frequency domain, with the amplitude (sign) of their elements being the amplitude (relative $\pi$ phase shift) of the sinusoidal signals at the fundamental frequency band. A constant frequency spacing d$f$ between neighboring sinusoidal signals is adjusted to fit all the elements into the working frequency range (i.e., bandwidth of our device).

The second-order nonlinear process of our device convolutes $\vec{a}$ and $\vec{b}$ and produces output vectors $\vec{A}$, $\vec{B}$, and $\vec{C}$ at the second-order frequency band. Specifically, vectors $\vec{A}$ and $\vec{B}$ are the self convolution of input vectors $\vec{a}$ and $\vec{b}$, respectively. Vector $\vec{C}$ is the cross convolution of input vectors $\vec{a}$ and $\vec{b}$, which is given by,

$$C_k = \mu \sum_{j=0}^{k} a_j b_{k-j}. \qquad (1)$$

where $\mu$ is the nonlinear conversion efficiency, $k$ is in range of 0 to $N + M - 2$. $N$ and $M$ are the lengths of vectors $\vec{a}$ and $\vec{b}$, respectively. From the perspective of wave physics, $C_k$ is the interference of all nonlinearly generated signals by $a_j$ and $b_{k-j}$ at frequency $f_{C_k} = f_{a_j} + f_{b_{k-j}} = f_{a_0} + f_{b_0} + k \mathrm{d}f$.

## III. Characterizations of Acoustic-wave Nonlinearity

To evaluate the acoustic-wave nonlinearity of our device, the input-output relationship and bandwidth of the second harmonic generation (SHG) are experimentally characterized. A 1,023.5-MHz sinusoidal signal at 0 dBm, which is at the designed fundamental IDT frequency, is applied at Port 1 to excite SAW. A spectrum analyzer (Keysight P5004A) is used to detect the output amplitude at the fundamental (second-order) frequency signal at

Port 2 (Port 3). Fig. 4(a) shows the measured fundamental frequency signal at 1,023.5 MHz and the SHG signal at 2,047.0 MHz, which is exactly double of the input fundamental frequency. The bandwidth of the SHG is measured by varying the input frequencies from 1,020 to 1,030 MHz with a constant power level of 0 dBm. The corresponding output power of SHG (Fig. 4(b)) shows a peak signal of -61.57 dBm at 1,023.5 MHz, with a 3 dB bandwidth of approximately 0.90 MHz. The fundamental IDT conversion efficiency is ~10% estimated from the measured transmission of an IDT pair. The nonlinear conversion efficiency at the peak frequency is approximately 7%/W on chip and 0.07%/W cable to cable. The efficiency is primarily influenced by the nonlinear coefficient of LN. By fixing the input frequency at 1,023.5 MHz and varying the input power, the output SHG power quadratically increases with the input power (Fig. 4(c)). Specifically, the measured input fundamental wave power and output second harmonic power hold a linear relationship on a logarithmic scale (with a slope of approximately 2) and a quadratic relationship on a linear scale.

## IV. Matrix-matrix Multiplication

Fig. 5 shows the scalability of matrix-matrix multiplications using our computing device. A microwave signal generator (Rigol DSG836A) with IQ modulation is used to generate the input signals. As an example, randomly generated 4×4 matrices $U$ ($V$) shown in Fig. 5(b) is simultaneously encoded in the fundamental frequency band row by row (column by column) with a frequency spacing $df$ of 100 Hz at a power level of 0 dBm (Fig. 5(a)). The multiplication result $W = UV$ is retrieved from the yellow data points (i.e., $C_{4k-1}$, $k = 1, 2, 3, …, 16$) within the cross convolution at the second-order frequency band. The measured $W$ shows good agreement with the expected $W$ (Fig. 5(c)). By reducing the frequency spacing to 20 Hz and maintaining the same power level, our device can encode two 16×16 randomly generated input matrices. The measured $W$ (Fig. 5(d)) and the expected $W$ (Fig. 5(e)) agree with each other, though with a slightly larger error than that of 4×4 matrices. Normalized root-mean-square error (NRMSE) is used to quantify the error between the measured $\widetilde{W}$ and the expected $W$:

$$\text{NRMSE} = 1 - \frac{(\Sigma_i \overline{W}_i W_i)^2}{\Sigma_i \overline{W}_i^2 \cdot \Sigma_i W_i^2}. \quad (2)$$

Fig. 5(f) shows NRMSE of measured $W$ when input matrices are two $N \times N$ random matrices, $N = 4, 8$, and 16. For each $N$, 50 independent cases are measured. As the matrix size increases from 4 to 16, NRMSE increases from an average of $10^{-5}$ to an average of $5 \times 10^{-3}$. This is mainly due to the increased low-frequency leakage for more IQ modulation frequencies, as indicated by the unexpected peak at $df$.

## V. Image Processing

Our device is versatilely programmable for image processing with various kernels. Fig. 6(a) illustrates an example of convolving a 28-by-28-pixel image from Fashion MNSIT with a 'Sobel XY' kernel. The gray value of image pixels and kernel elements are encoded into the fundamental frequency band row-by-row with a frequency spacing $df$ of 100 Hz at a power level of 0 dBm. In the second-order frequency band, the output image is retrieved from the cross convolution. As shown in Fig. 6(b), five different image kernels (i.e., blur, sharpening, Sobel X, Sobel Y, and Sobel XY) are tested. Compared with the calculated expected output on the left, the mean NRMSE of the five measured output images is $1.3 \times 10^{-3}$. By reducing the frequency spacing to 20 Hz and increasing the power level to 20 dBm, our device can perform edge detections on a 128-by-128-pixel image. The image is separately convoluted with two 3×3 kernels to highlight horizontal and vertical edges, as shown in Fig. 6(c). Here, the kernel for horizontal edge detection is [-1, -1, -1; 0, 0, 0; 1, 1, 1], while the kernel for vertical edge detection is [-1, 0, 1; -1, 0, 1; -1, 0, 1]. The two output images are then combined to show the full edge detections. The measured outputs agree with the expected outputs on the left. In terms of the number of MAC operations associated with the edge detection, 0.0144 tera floating-point operations per second (TFLOPS) is performed in the single device, leading to a computing area density of 0.48 TFLOPS/mm$^2$ and a core power efficiency of 0.14 TFLOPS/Watt.

## VI. Summary

This work presents a scalable, programmable, frequency-domain parallel computing paradigm leveraging gigahertz (GHz)-frequency acoustic-wave nonlinearities. We demonstrate multiplications of matrices and convolutional image processing of photos. Further improvement in the computing performance (e.g., speed, computing area density, and power efficiency) is possible by increasing the working frequency and the bandwidth of our device. Our technology could find versatile applications in near-sensor signal processing and edge computing.

Acknowledgment

Device fabrication was conducted as part of a user project (CNMS2022-B-01473) at the Center for Nanophase Materials Sciences (CNMS), which is a US Department of Energy, Office of Science User Facility at Oak Ridge National Laboratory. Research was sponsored by the Air Force Office of Scientific Research (AFOSR) and was accomplished under Grant Number W911NF-23-1-0235. The views and conclusions contained in this document are those of the authors and should not be interpreted as representing the official policies, either expressed or implied, of the Army Research Office or the U.S. Government. The U.S. Government is authorized to reproduce and distribute reprints for Government purposes notwithstanding any copyright notation herein.

## Device design and computing principles

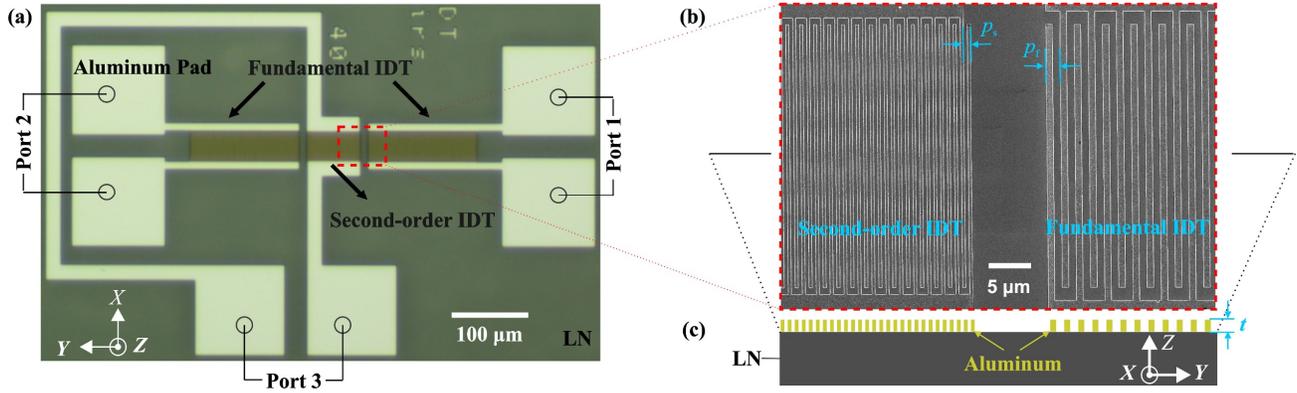

Fig. 1. (a) Optical microscopic image of nonlinear acoustic-wave device. The device is on a Z-cut lithium niobate (LN) substrate, with the nonlinear acoustic waves propagating along the crystal Y-axis. Only a layer of patterned 100-nm-thick aluminum is deposited on top of LN to form the interdigital transducers (IDTs). IDTs on the sides (Port 1 and Port 2) are used to excite and detect acoustic waves at the fundamental frequency band, while the middle IDT (Port 3) is for acoustic waves at the second-order frequency band. Bright regions are aluminum and dark green regions are LN. (b) A close-up view of the central region where the acoustic-wave nonlinearity is leveraged for computing. The pitch of fundamental (second-order) IDT fingers $p_f$ ($p_s$) is 1.84 (0.908) μm, which corresponds to the half wavelength of SAW at the fundamental (second-order) frequency. (c) Cross-section view of the fundamental and second-order IDTs. The thickness of aluminum $t$ is 100 nm.

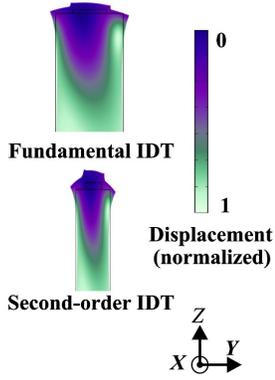

Fig. 2. SAW mode of the fundamental (second-order) IDT at 1,047 (2,093) MHz in simulation, with the color scale representing the normalized displacement field.

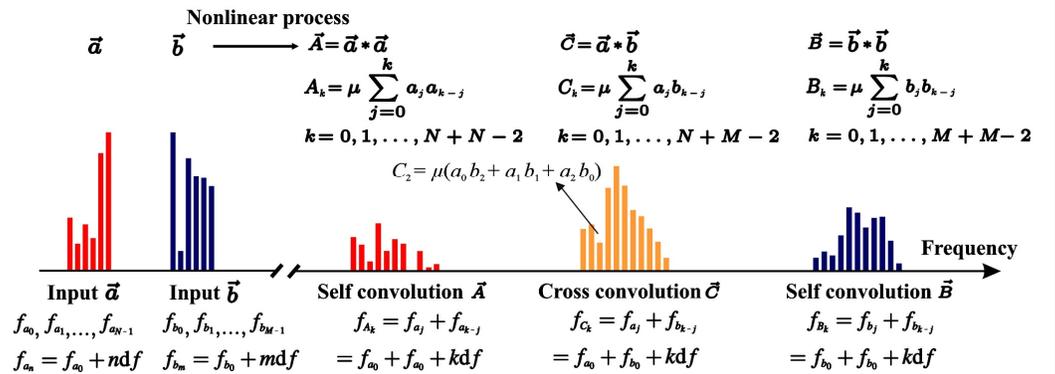

Fig. 3. Nonlinear computing principle. Floating-point input vectors $\vec{a}$ and $\vec{b}$ are encoded in the frequency domain via Port 1, with the amplitude (sign) of their elements being the amplitude (relative π phase shift) of the sinusoidal signals at the fundamental frequency band, respectively. Benefiting from the acoustic nonlinearity, self (cross) convolution $\vec{A}$ and $\vec{B}$ ($\vec{C}$) at the second-order frequency band can be retrieved at Port 3 for computing applications such as matrix-matrix multiplications and image processing.

## Characterizations of acoustic-wave nonlinearity on device

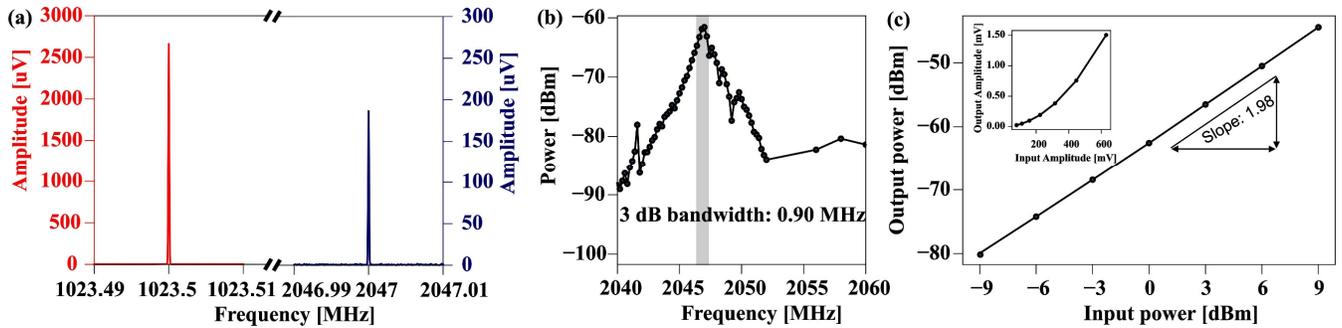

Fig. 4. Characterization of acoustic-wave nonlinearity via second harmonic generation (SHG). (a) The measured amplitude of fundamental (second-order) signals, when a 1,023.5-MHz signal at 0 dBm is generated at Port 1 via a signal generator and the transmitted signal amplitude at Port 2 ( Port 3) is measured by a spectrum analyzer. The generated SHG signal is at 2,047 MHz. (b) Bandwidth of SHG. The output powers of SHG at different input frequencies. The input signal power is 0 dBm. (c) The input-output power relationship of SHG. The input frequency is at 1,023.5 MHz Inset: the input-output amplitude relationship in linear scale.

**Matrix-matrix multiplications**

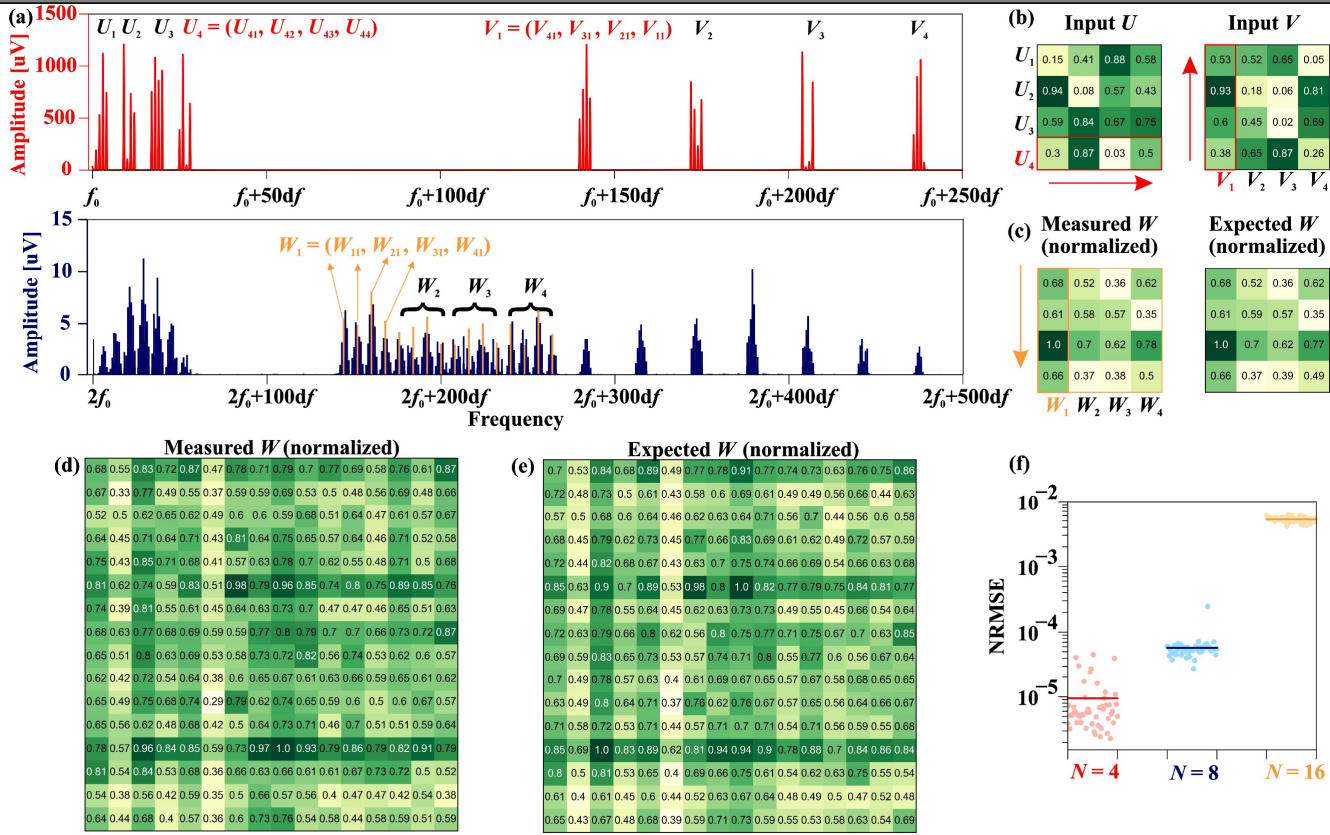

Fig. 5. (a) Randomly generated 4×4 Matrix $U$ ($V$) is encoded in the fundamental frequency band row by row (column by column) with $df = 100$ Hz. The multiplication result $W = UV$ is retrieved from the yellow data points (i.e., $C_{4k-1}$, $k = 1, 2, 3, …, 16$) within the cross convolution in the second-order frequency band. (b) The input matrices $U$ and $V$. (c) The measured and the expected $W$ for inputs in (b). (d) The measured $W$ and (e) the expected $W$ for the multiplication of two randomly generated matrices $U$ and $V$ at a size of 16×16. (f) Normalized root-mean-square error (NRMSE) between the measured and the expected $W$ for $N = 4, 8,$ and 16.

**Image processing**

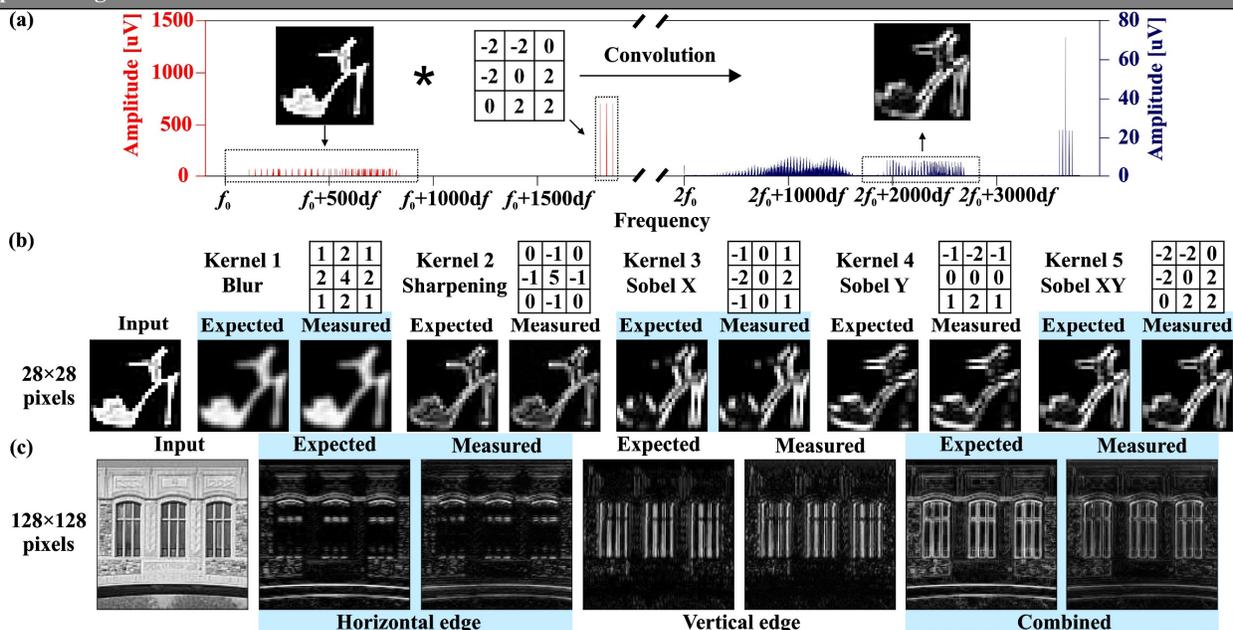

Fig. 6. (a) Both the image pixels (28×28) and kernel elements (3×3) are encoded in the fundamental frequency band row by row with $df = 100$ Hz. The amplitude of the signal represents the grayscale of the pixel. The output image is retrieved from the cross convolution in the second-order frequency band. (b) The 28-by-28-pixel image is convoluted with five different kernels using our device. (c) A 128-by-128-pixel image is convoluted with two 3×3 kernels to highlight horizontal/vertical edges, which is then combined to show edge highlighting. The measured images after convolution agree with the expected images.